\documentclass{Interspeech2024}
\usepackage{xcolor}
\usepackage{color}
\usepackage{graphicx}
\usepackage{subcaption}
\usepackage{multirow}
\usepackage{cite}

\interspeechcameraready

\title{LUPET: Incorporating Hierarchical Information Path into Multilingual ASR}
\name[affiliation={1}]{Wei}{Liu$^{*,}$}
\name[affiliation={2}]{Jingyong}{Hou}
\name[affiliation={2}]{Dong}{Yang}
\name[affiliation={2}]{Muyong}{Cao}
\name[affiliation={1}]{Tan}{Lee}
\address{
  $^1$ Department of Electronic Engineering, The Chinese University of Hong Kong \\
  $^2$ GVoice, Tencent
  }
\email{
\thanks{$^{*}$ This work was done during an internship at Tencent.}
louislau\_1129@link.cuhk.edu.hk, \{jingyonghou,daviddyang,locwellcao\}@tencent.com, tanlee@cuhk.edu.hk}

\begin{document}
\maketitle  
\begin{abstract}

Toward high-performance multilingual automatic speech recognition (ASR), various types of linguistic information and model design have demonstrated their effectiveness independently. They include language identity (LID), phoneme information, language-specific processing modules, and cross-lingual self-supervised speech representation. It is expected that leveraging their benefits synergistically in a unified solution would further improve the overall system performance. This paper presents a novel design of a hierarchical information path, named LUPET, which sequentially encodes, from the shallow layers to deep layers, multiple aspects of linguistic and acoustic information at diverse granularity scales. The path starts from LID prediction, followed by acoustic unit discovery, phoneme sharing, and finally token recognition routed by a mixture-of-expert. ASR experiments are carried out on 10 languages in the Common Voice corpus. The results demonstrate the superior performance of LUPET as compared to the baseline systems. Most importantly, LUPET effectively mitigates the issue of performance compromise of high-resource languages with low-resource ones in the multilingual setting.

\end{abstract}

\section{Introduction}

Conventionally an automatic speech recognition (ASR) system is developed to transcribe speech into text for a specific language. Toward multilingual ASR, recent research is focused on building a unified model that covers multiple languages~\cite{li2021scaling, radford2023robust, zhang2023google, DBLP:conf/interspeech/PratapSTHLSC20}. One practical advantage is the reduction of training and deployment costs, as compared to building a separate monolingual model for each language. It also facilitates sharing of linguistic knowledge among the languages and may help elevate the recognition performance on those with limited data resources~\cite{DBLP:conf/lrec/YadavS22}. It was shown that training a fully shared end-to-end (E2E) model on a multilingual speech corpus is a simple yet effective solution (\textit{vanilla})~\cite{DBLP:conf/interspeech/PratapSTHLSC20}. The multilingual corpus is made by mixing the corpora of different languages, and a shared vocabulary is used. Due to the heterogeneous nature of different languages~\cite{li2022massively, DBLP:conf/interspeech/PratapSTHLSC20}, the \textit{vanilla} scheme exhibits the issue that the recognition performance on the high-resource languages inevitably compromises in the multilingual training, in order to attain reasonable performance on the low-resource languages.


To mitigate the issue of performance compromise, there have been attempts that incorporate language identity (LID) information~\cite{watanabe2017language, DBLP:conf/interspeech/ZhangLSSMCH22, zhou2022configurable,chen2023improving}, phoneme information~\cite{sailor2020multilingual, zhu2021multilingual, wang2021unispeech}, and language-specific architecture, e.g., mixture-of-expert (MoE)~\cite{gaur2021mixture, you2022speechmoe2,  hu2023mixture, wang2023language, DBLP:conf/asru/SunLHZZXWLLLG23}. These approaches typically require a supervised training process, which requires labeled training data. On the other hand, self-supervised learning (SSL) is believed to be an effective way of cross-lingual data sharing. The representative works include wav2vec2.0~\cite{baevski2020wav2vec}, HuBERT~\cite{hsu2021hubert}, XLSR~\cite{DBLP:conf/interspeech/ConneauBCMA21}, and many others. The two-stage training scheme, i.e., pre-training and fine-tuning, has been widely applied in SSL. In~\cite{bai2022joint}, joint unsupervised and supervised training (JUST) was shown to outperform two-stage training on multilingual ASR. 
JUST uses a contrastive loss and a masked language model (MLM) loss to learn discrete units for better contextualized representations.

In the present study, a hierarchical information path is developed to combine multiple useful factors synergistically to boost the overall performance of a multilingual ASR system. The path comprises a sequence of prediction modules that incorporate linguistic and acoustic information at diverse granularity levels into the recognition process. These modules are namely, \textbf{L}ID, Acoustic \textbf{U}nit discovery, \textbf{P}honeme sharing, and mixture of \textbf{E}xperts for \textbf{T}oken recognition. We use the acronym \textbf{LUPET} to denote the proposed design, in which each alphabet represents one of the information components in the path. The path LUPET can be easily integrated into a vanilla ASR architecture by unfolding with the encoder layers. Within this path, information from a shallow layer can benefit those that occur in deeper ones, and hence it is considered a hierarchical flow. Here the shallow layer refers to an encoder layer close to the input. Importantly, the required information labels are either straightforward to obtain or can be derived via an SSL process.



The effectiveness of LUPET is evaluated by experiments on 10 languages in the Common Voice~\cite{DBLP:conf/lrec/ArdilaBDKMHMSTW20}. The results show that, compared to the vanilla system, LUPET can achieve 19.7\% and 12.3\% relative reduction of average word error rate with CTC and attention decoding, respectively. LUPET also outperforms previous baseline systems. In particular, it demonstrates superior performance on the high-resource languages as its performance compromise to low-resource languages is alleviated.

\begin{figure}[t!]
\vspace{-10pt}
  \centering
  \includegraphics[width=0.83\linewidth]{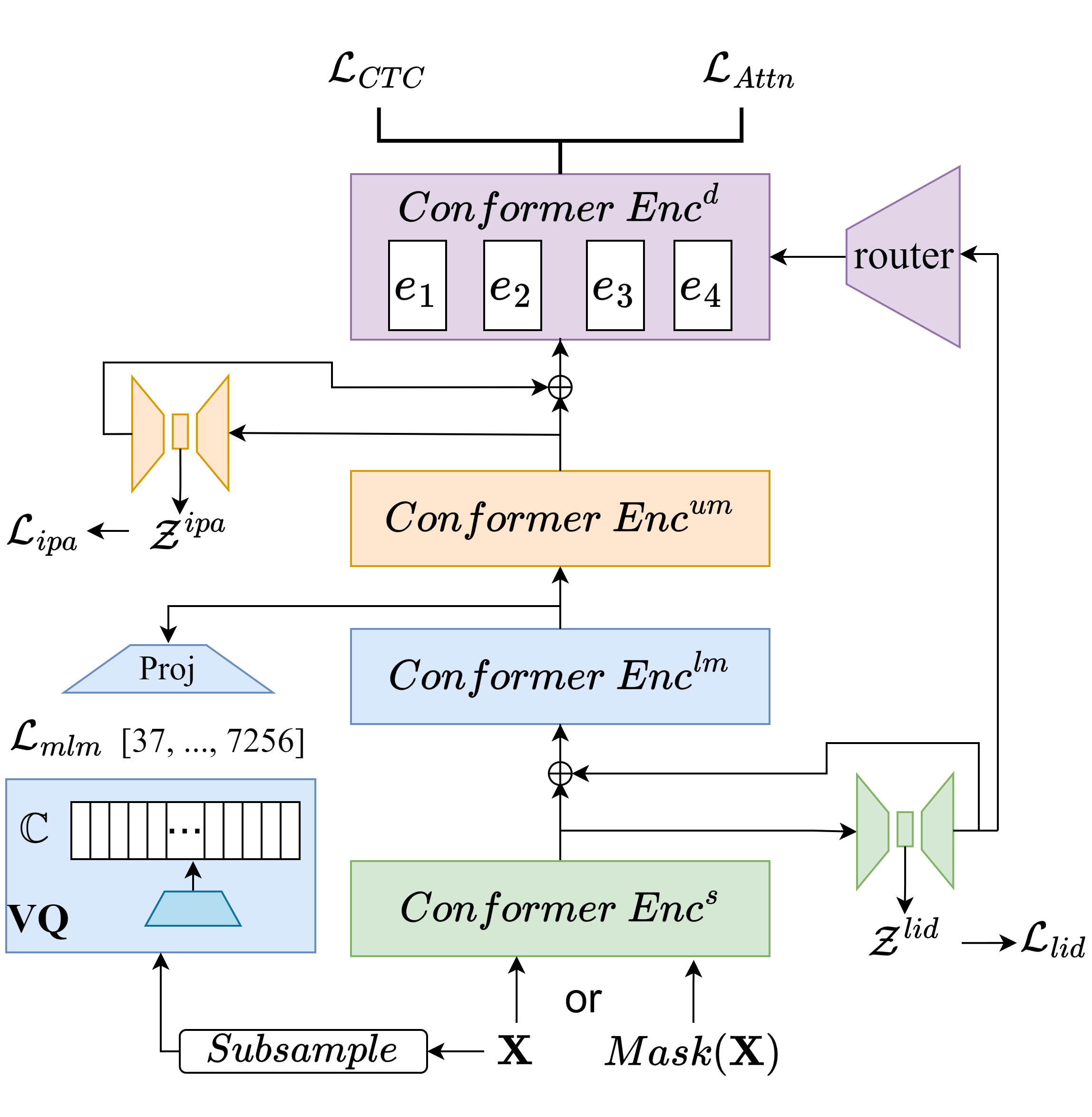}
  \vspace{-14pt}
  \caption{The overall architecture of our proposed LUPET multilingual ASR. LUPET information path unfolds with the encoder layers. \{ $Enc^{s}$, $Enc^{lm}$, $Enc^{um}$, $Enc^{d}$\} represent \textit{shallow}, \textit{lower-middle}, \textit{upper-middle}, \textit{deep} layers, respectively. $Enc^{s}$ and  $Enc^{um}$ are used for LID and IPA phoneme prediction. $Enc^{lm}$ performs acoustic unit discovery with a random-projection quantizer, where $\mathbb{C}$ denotes the codebook for vector quantization (VQ). $Enc^{d}$ denotes conformer layers modified with MoE which consists of four experts and a router. All trapezoid modules refer to linear projection. }
  \label{fig:lppet}
  \vspace{-5mm}
\end{figure}

\section{LUPET}

\subsection{Vanilla E2E Multilingual ASR}
\label{subsec: vanilla ASR}
The E2E multilingual ASR architecture we adopt is the hybrid CTC-Attention conformer~\cite{watanabe2017hybrid, DBLP:conf/interspeech/GulatiQCPZYHWZW20}. It consists of three components, namely encoder, decoder, and CTC~\cite{graves2006connectionist} layer. The encoder takes an acoustic feature sequence $\mathbf{X}=\{\mathbf{x}_{t}\}_{t=1}^{T}$ as input and converts it to hidden representation $\mathbf{H} = \{\mathbf{h}_t\}_{t=1}^{T^{'}}$, where $T$ and $T^{'}$ denote the number of original frames and the sub-sampled frames. 
The $\mathbf{H}$ is then forwarded to two classification branches for predicting the token sequence $\mathbf{Y} = \{y_u \in \mathcal{V}\}_{u=1}^{U}$, where $\mathcal{V}$ is a shared multilingual vocabulary built by BPE~\cite{DBLP:conf/acl/ZouharMGDVSC23} and $U$ denotes the number of tokens. One classification branch, i.e., decoder, conditions $\mathbf{H}$ to autoregressively compute token-level posterior $p(y_u|\mathbf{H}, y_{1:u-1})$ via a cross attention mechanism. The attention loss is given as \vspace{-5pt}
\begin{equation}
\vspace{-2pt}
\label{eq:loss_attn}
    \mathcal{L}_{Attn} = -\sum_{u=1}^{U} log p(y_u|\mathbf{H}, y_{1:u-1}).
    \vspace{-2mm}
\end{equation}
Another classification branch, i.e., the CTC layer, simultaneously derives the frame-level posteriors $p(\mathbf{z}_t|\mathbf{h}_t)$. The CTC loss is formulated as follows: \vspace{-8pt}
\begin{equation}
\label{eq:loss_ctc}
    \mathcal{L}_{CTC} = CTC(\mathbf{Z}, \mathbf{Y}) = -\sum_{Z\in B^{-1}(\mathbf{Y})} \sum_{t=1}^{T^{'}} logp(\mathbf{z}_t|\mathbf{h}_t),
\end{equation}
where $\mathbf{z}_t$ is the logits over $\mathcal{V} \cup \emptyset$ and $B^{-1}$ is the inverse function that gives all valid alignment paths between input sequence $\mathbf{H}$ and output sequence $\mathbf{Y}$. The blank token $\emptyset$ is specially designed by CTC for aligning $\mathbf{H}$ and $\mathbf{Y}$. $\mathbf{Z} = \{\mathbf{z}_{t}\}_{t=1}^{T^{'}}$ represents one possible alignment path. The training objective of hybrid CTC-Attention is a linear combination of Eq. \ref{eq:loss_attn} and Eq. \ref{eq:loss_ctc}:
\begin{equation}
\label{eq:loss_ctc_attn}
    \mathcal{L}_{CTC-Attn} = (1 - \lambda) \mathcal{L}_{Attn} + \lambda \mathcal{L}_{CTC},
\end{equation}
where $\lambda$ is a coefficient to control the weight of CTC loss.

\subsection{Incorporating LUPET}
Recent studies have pointed out (1) Both LID and phoneme information are beneficial for multilingual training~\cite{chen2023improving, wang2021unispeech}; (2) Design language-specific modules to process language-specific information is useful to reduce language interference~\cite{hu2023mixture, DBLP:conf/asru/SunLHZZXWLLLG23}; 
(3) The success of self-supervised cross-lingual representation learning applied in ASR \cite{DBLP:conf/interspeech/ConneauBCMA21, chiu2022self}. Although these factors' effectiveness has been verified separately, how to synergistically combine them to contribute a better solution from a unified perspective remains an open question. 
The proposed LUPET provides a novel view from the multilingual hierarchical information path. From LID to acoustic unit followed by phoneme then go through MoE routing to the final token, the information that occurred in the early position of the path is assumed to contribute to the prediction of later information of the path.  

As shown in Fig.~\ref{fig:lppet}, the full encoder is composed of \{ $Enc^{s}$, $Enc^{lm}$, $Enc^{um}$, $Enc^{d}$\}, from shallow layers to deep layers. LUPET information path unfolds with the encoder layers. The shallow layers of encoder $Enc^{s}$ are used to identify the spoken language. Denote the output of  
$Enc^{s}$ as shallow representations $\mathbf{H}^{s}$.  $\mathbf{H}^{s}$ is then projected to the LID logits $\mathbf{Z}^{lid}$ via a linear transformation. The logits dimension $dim(\mathbf{Z}^{lid})=$ \#$LID + 1$, where $1$ represents a special blank token for CTC as mentioned in Sec.~\ref{subsec: vanilla ASR}.
The LID prediction loss is formulated as:
\begin{equation}
\label{eq:loss_lid}
    \mathcal{L}_{lid} = CTC(\mathbf{Z}^{lid}, LID_{seq}),
\end{equation}
where the sequential LID labels $LID_{seq}$ are constructed by repeating the single LID label to the number of output tokens. 

The predicted LID information then is propagated to subsequent layers of the encoder via self-conditioning
\begin{equation}
\label{eq:lid_sc}
    \mathbf{H}^{s'} = \mathbf{H}^{s} + LIN(\mathbf{Z}^{lid}),
\end{equation}
where $LIN$ denotes a linear layer to keep the hidden dimension and $\mathbf{H}^{s'}$ is the input representations of $Enc^{lm}$.

The lower-middle layers of encoder $Enc^{lm}$ are utilized to perform acoustic unit discovery. Similar to BEST-RQ \cite{chiu2022self}, a random-projection quantizer including a projection matrix $Proj^{c}$ and codebook $\mathbb{C}$ is applied and none of the parameters are trainable. Vector quantization (VQ) is carried out on acoustic features $\mathbf{X}$ to produce discrete labels 
\begin{equation}
    \mathbf{Lab}_{u} = \mathbb{C}(Proj^{c}(Sub(\mathbf{X})))), 
\end{equation}
where $Sub$ represents the subsample operation and $Proj^{c}$ performs projection from the speech feature dimension to the code vector dimension. The output of $\mathbb{C}$ are the indices of the nearest code vectors of the codebook to the input vectors.

With probability $p$, $\mathbf{X}$ is randomly masked to feed the encoder. Masked language modeling (MLM) is then performed to predict $\mathbf{Lab}_{u}$. Denote $\mathbf{H}^{lm}_{M}$ as the output representation of $Enc^{lm}$, where the under-script $M$ means having Mask($\mathbf{X}$) as input.  Let $\mathbf{mi}$ be the masked indices on $\mathbf{H}^{lm}_{M}$, the MLM loss can be written as:
\begin{align}
    \label{eq:loss_mlm}
    \mathcal{L}_{mlm} = CE(Proj^{u}(\mathbf{H}^{lm}_{M}[\mathbf{mi}]), \mathbf{Lab}_{u}[\mathbf{mi}]),
\end{align}
where $Proj^{u}$ projects the hidden dimension to the size of the codebook and MLM loss is the cross-entropy ($CE$) between logits over the codebook and labels at the masked positions.  
 
Discrete acoustic units discovered by $Enc^{lm}$ are expected to facilitate pronunciation learning to incorporate phonetic information for subsequent layers. Similar to LID prediction, the output representation $\mathbf{H}^{um}$ of upper-middle encoder $Enc^{um}$ is used to predict phoneme sequence $IPA$. $\mathbf{Z}^{ipa}$, projected by $\mathbf{H}^{um}$, is the logits over IPA phonemes and an additional blank token. Eq. \ref{eq:loss_ipa} gives the loss of IPA prediction: 
\begin{equation}
\label{eq:loss_ipa}
     \mathcal{L}_{ipa} = CTC(\mathbf{Z}^{ipa}, IPA).
\end{equation}
Following Eq. \ref{eq:lid_sc}, self-conditioning is similarly applied to obtain $\mathbf{H}^{um'}$ to propagate the predicted phonetic information.
\begin{equation}
\label{eq:ipa_sc}
    \mathbf{H}^{um'} = \mathbf{H}^{um} + LIN(\mathbf{Z}^{ipa})
\end{equation}
Lastly, the deep layers of encoder $Enc^{d}$ are modified with MoE. Multiple FFN experts and a routing network are included in the MoE structure. The language self-condition representation ($LIN(\mathbf{Z}^{lid})$ in Eq. \ref{eq:lid_sc}) is regarded as the LID embedding to feed the routing network. The output of the routing network is a softmax distribution over the number of experts.  Followed \cite{hu2023mixture}, the top-2 experts with the highest probabilities are dynamically routed to process each frame based on the frame-level LID information from shallow encoder layers. 

To incorporate the hierarchical information, from LID, acoustic unit, phoneme, and token,  the objective function of LUPET is given as a linear combination of Eq. (\ref{eq:loss_ctc_attn}, \ref{eq:loss_lid}, \ref{eq:loss_mlm}, \ref{eq:loss_ipa}):
\begin{equation}
\label{eq:loss_lupet}
    \mathcal{L}_{LUPET} = \mathcal{L}_{CTC-Attn} + w_{1} \mathcal{L}_{lid} + w_{2} \mathcal{L}_{mlm}+ w_{3} \mathcal{L}_{ipa},
\end{equation}
where $w_1, w_2, w_3$ are the weights of the corresponding losses.

\begin{table}[t!]
\centering
\caption{Training and testing hours of 10 languages from the Common Voice 13.0 corpus in our experiments.} \vspace{-10pt}
\resizebox{0.8\linewidth}{!}{
\begin{tabular}{c|ccccc}
\toprule
LID   & en      & fr     & es     & zh     & it     \\ \midrule
Train & 2279.98 & 872.19 & 448.45 & 359.91 & 286.61 \\
Test  & 26.90    & 26.08  & 26.65  & 30.44  & 26.27  \\ \midrule
LID   & ru      & pt     & tr     & nl     & tt     \\ \midrule
Train & 178.78  & 125.35 & 69.08  & 73.82  & 19.95  \\
Test  & 15.73   & 11.91  & 12.05  & 14.58  & 5.70    \\ \bottomrule
\end{tabular}}
\label{tab:dataset}
\vspace{-12pt}
\end{table}

\section{Experimental Setup}
\subsection{Dataset}
The 10 languages, namely English (en), French (fr), Spanish (es), Chinese (zh), Italian (it), Russian (ru), Portuguese (pt), Turkish (tr), Dutch (nl) and Tatar (tt) from the public available Common Voice 13.0 \cite{DBLP:conf/lrec/ArdilaBDKMHMSTW20} are selected for our multilingual ASR experiments. The language coverage includes high-resource languages, e.g., English with around 2,280 hours of training data, and low-resource languages, e.g., Tatar with only about 20 hours. The detailed training and testing statistics are listed in Tab.~\ref{tab:dataset}. Note that zh includes Mandarin, Taiwanese, and Cantonese. A standard text normalization (the same as in Whisper~\cite{radford2023robust}) is applied to all transcriptions of the dataset. 

\subsection{Multilingual ASR Configurations}
\subsubsection{Vanilla}
The vanilla model adopts a hybrid CTC-Attention architecture. The encoder has 12 conformer layers with 8 attention heads and 512 hidden dimensions, while the decoder has 6 transformer layers~\cite{DBLP:conf/nips/VaswaniSPUJGKP17}. The CTC weight $\lambda$ in Eq.~\ref{eq:loss_ctc_attn} is set to $0.3$. The input acoustic feature to the network is the typical 80-dimensional log-Mel filterbank. The output vocabulary used is derived from Whisper's tokenizer. This tokenizer was obtained by BPE using UTF-8 bytes of the entire training dataset of Whisper.

\subsubsection{LUPET}
Compared to \textit{vanilla}, the encoder architecture has several modifications by incorporating LUPET.  The output positions of \{ $Enc^{s}$, $Enc^{lm}$, $Enc^{um}$, $Enc^{d}$\} are at the \{3-th,6-th,9-th,12-th\} layer of the original encoder, respectively. When MLM loss takes effect, acoustic feature $\mathbf{X}$ is randomly masked consecutive 20 frames with probability $p=0.01$. 
Codebook $\mathbb{C}$ of the random-projection quantizer has a size of 8192 and a dimension of 16. IPA sequence per-utterance is obtained using an open-sourced toolkit \textit{phonemizer}~\cite{Bernard2021}. In each layer of $Enc^{d}$, MoE including 8 FFN experts is used to replace the end-FFN of the original conformer layer. In Eq.~\ref{eq:loss_lupet}, the weight coefficients $w_1$, $w_2$, and $w_3$ are set to 0.3, 0.07, and 0.3, respectively.

\subsubsection{Baselines}
Several baselines are used for comparison: (1) \textit{Mono}, monolingual ASR with vanilla architecture and 256 hidden dimensions is trained per language. (2) \textit{Oracle\_LID}, append the pre-known LID embedding to input acoustic feature for multilingual ASR training. (3) \textit{MoE}~\cite{hu2023mixture}, keep the $Enc^{d}$ of LUPET and remove other auxiliary losses, hidden representation $\mathbf{H}^{um}$ is used as the input to the routing network. (4) \textit{LID\_SC}~\cite{chen2023improving}, LID prediction by CTC and LID information self-conditioning (SC) are performed over the vanilla model. (5) \textit{Whisper}, whisper-large-v2 with oracle LID is used for decoding. 

\subsection{Training Scheme and Evaluation Metric}
We implement the vanilla and our proposed LUPET methods on the Wenet toolkit~\cite{DBLP:conf/interspeech/YaoWWZYYPCXL21}. 
The model is trained with Adam~\cite{DBLP:journals/corr/KingmaB14} optimizer with a learning rate (LR) of $1e-3$. LR schedule has a warmup step of $15000$. Batch size is set to $12$ with $accum\_grad=16$. 8 V100 GPUs are used for DDP training. Each multilingual model is trained for 50 epochs and each monolingual model is trained for 100 epochs. If not specified otherwise, MLM takes effect from epoch 5 to 30. The final model for decoding is obtained by averaging the $10$ best models with the lowest validation losses. Character error rate (CER) for Chinese and word error rate (WER) for other languages are used to measure the system's performance.

\section{Results and Analysis}
\begin{table*}[t!]
\vspace{-3mm}
\centering
\caption{WER (\%) results of different systems on 10 languages of Common Voice by CTC greedy decoding. \textit{avg 5high} denotes the averaged WER results of top-5 high-resourced languages and \textit{avg 5low} is similar for the low-resourced case. In the LUPET block, the \textbf{backslash /\ } represents the ablation study by removing the following component, where U = acoustic unit discovery, P = IPA sharing, and L = LID prediction. $w_2=1$ means the weight coefficient of $\mathcal{L}_{mlm}$ is set as 1. Uto50ep means U takes effect until 50 epochs. }
\vspace{-5pt}
\resizebox{0.95\linewidth}{!}{
\begin{tabular}{l|cccccccccc|cccc}
\toprule
Model       & en    & fr    & es   & zh    & it    & ru    & pt    & tr    & nl    & tt     & avg   & avg w/o tt & avg 5high & avg 5low \\ \midrule
Mono        & 13.03 & 12.51 & 9.37 & 13.01 & 11.15 & 11.55 & 11.16 & 25.73 & 19.34 & 83.62  & 21.05 & 14.09      & 11.81     & 30.28    \\
Vanilla     & 13.50  & 13.33 & 9.74 & 13.71 & 10.56 & 16.34 & 12.07 & 23.49 & 13.72 & 36.78  & 16.32 & 14.05      & 12.17     & 20.48    \\
Oracle\_LID & 12.69 & 12.08 & 8.51 & 12.8  & 9.07  & 13.64 & 9.39  & 20.12 & 11.72 & 30.29  & 14.03 & 12.22      & 11.03     & 17.03    \\
LID\_SC     & 12.94  & 12.24 & 8.84 & 13.46 & 9.36  & 14.72 & 10.89 & 22.53 & 12.74 & 34.19  & 15.19 & 13.08      & 11.37     & 19.01    \\
MoE         & 12.86 & 12.81 & 9.23 & 12.67 & 9.91  & 13.56 & 10.26 & 20.55 & 11.46 & 30.47  & 14.38 & 12.59      & 11.50     & 17.26    \\ \midrule
LUPET       & \textbf{11.75} & \textbf{11.79} & \textbf{8.22} & \textbf{12.41} & \textbf{8.81}  & 10.95 & \textbf{8.95}  & \textbf{17.71} & 11.32 & 29.12  & \textbf{13.10} & \textbf{11.32}      & \textbf{10.60}     & 15.61    \\
LUPET / U         & 12.33 & 12.32 & 8.73 & 12.42 & 9.45  & \textbf{10.58} & 9.62  & 18.00    & \textbf{10.86} & 27.76  & 13.21 & 11.59      & 11.05     & \textbf{15.36}    \\
LUPET / P         & 12.35 & 12.22 & 8.67 & 12.31 & 9.39  & 11.92 & 10.51 & 21.92 & 12.08 & 27.23  & 13.86 & 12.37      & 10.99     & 16.73    \\
LUPET / UP        & 12.71 & 12.38 & 8.82 & 12.16 & 9.54  & 11.9  & 10.34 & 20.89 & 11.72 & \textbf{26.47}  & 13.69 & 12.27      & 11.12     & 16.26    \\
LUPET / LU        & 11.96 & 12.02 & 8.46 & 12.21 & 9.08  & 10.99 & 9.44  & 19.49 & 11.03 & 31.90   & 13.66 & 11.63      & 10.75     & 16.57    \\
LUPET $w_2 =1$        & 11.80  & 11.86 & 8.54 & 12.33 & 9.10   & 12.38 & 10.25 & 17.85 & 10.30  & 33.11  & 13.75 & 11.60      & 10.73     & 16.78    \\
LUPET Uto50ep     & 11.72 & 12.09 & 8.40  & 12.73 & 9.02  & 11.90  & 9.98  & 19.06 & 11.96 & 31.35  & 13.82 & 11.87      & 10.79     & 16.85    \\ \bottomrule
\end{tabular}}
\label{tab:wer_result}
\vspace{-12pt}
\end{table*}

\begin{figure}[t!]
\vspace{-2mm}
\vspace{-10pt}
  \centering
  \includegraphics[width=0.74\linewidth]{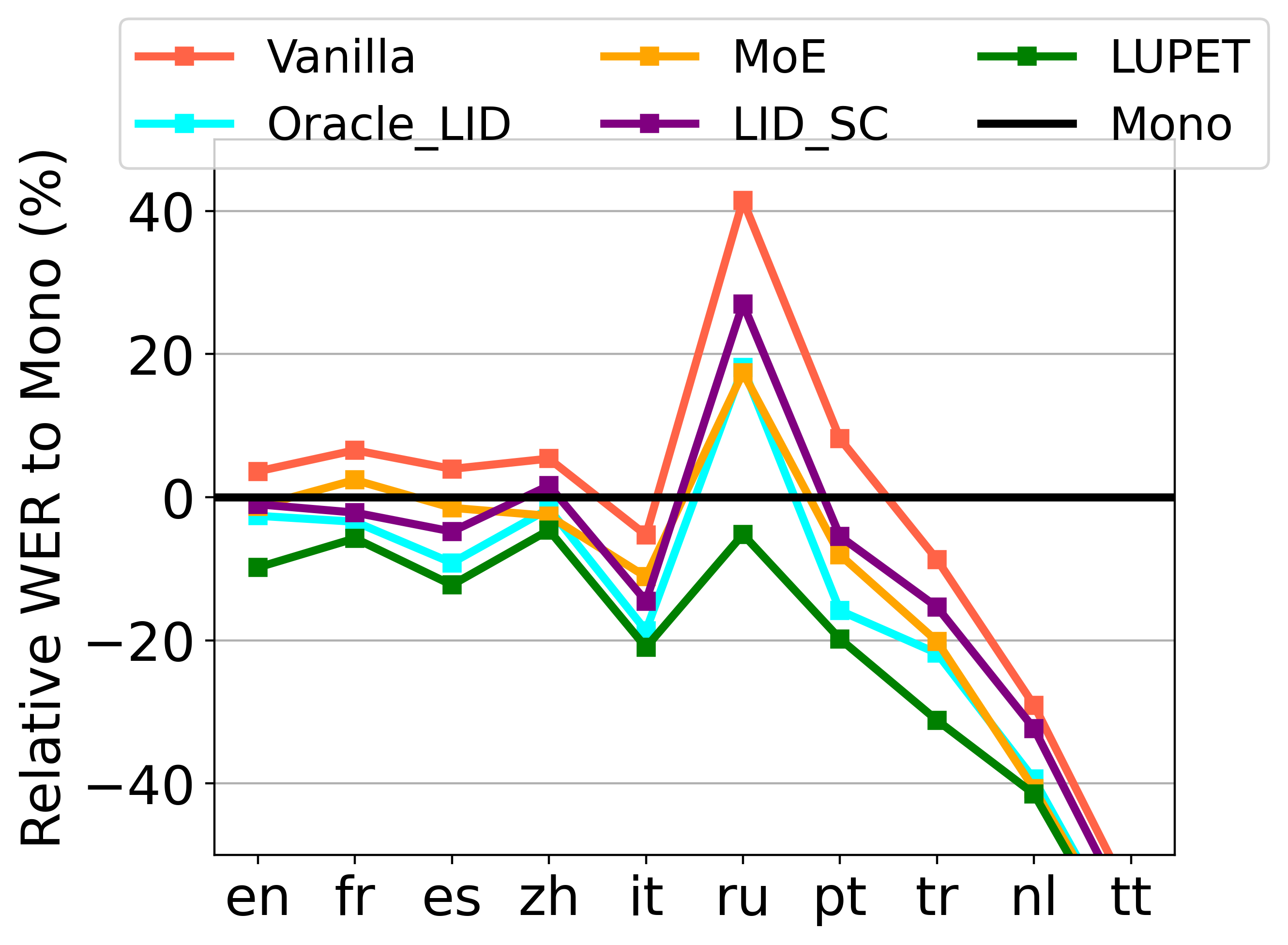}
  \vspace{-10pt}
  \caption{Relative WER changes of different systems to monolingual systems on 10 languages by CTC greedy decoding. }
  \label{fig:rwer}
  \vspace{-6mm}
\end{figure}

\subsection{Performance Comparison to Monolingual System}
A desirable multilingual ASR is expected to present better performance than its monolingual counterparts (\textit{Mono}).
As shown in Fig.~\ref{fig:rwer}, \textit{LUPET} and other baselines are used to compare performance to monolingual systems on the test sets of 10 languages. The x-axis follows an order from high-resource to low-resource languages. The y-axis denotes the relative WER to Mono, where the more negative value represents the lower WER. It is clear to see all curves are basically decreasing except the peak at ru. The decreasing trend is straightforward as the more low-resourced languages can achieve more significant performance gains. Multilingual training brings degradation to the Russian (ru) language in most cases. We speculate this may be due to the compromised phenomenon from Russian (ru) to Tarta (tt). The language tt achieves above 60\% relative WER reduction via multilingual training at the cost of side effects on language ru, belonging to a different language family with tt. 

It is worth noted that the recognition performance of \textit{Vanilla} system on four high-resource languages (en, fr, es, zh) cannot surpass the corresponding monolingual system, demonstrating the general compromised phenomenon towards low-resource languages in multilingual training. \textit{Oracle\_LID} system brings consistent improvements over \textit{Vanilla} across all languages, which proves the benefits of LID information. Both \textit{MoE} and \textit{LID\_SC} also show obviously better performance than \textit{Vanilla}, while being inferior to \textit{Oracle\_LID}. \textit{LUPET} outperforms all other baselines, serving as the only system that gives WER reduction on all languages compared to \textit{Mono}. The advantage of \textit{LUPET} is highlighted by the superior performance on high-resource languages, which largely mitigates the compromised phenomenon during multilingual training.   

\subsection{LUPET's Effectiveness Verification}
Tab. \ref{tab:wer_result} presents the WER results of different systems on 10 languages.  Averaged WER are calculated for overall comparison. The top-5 languages, with more training data, are roughly referred to as high-resources (\textit{5high}), while the remaining 5 languages are low-resources (\textit{5low}). Having similar observation from Fig.~\ref{fig:rwer}, \textit{LUPET} gives significantly better performance on all languages compared to other baselines. The benefits of LID prediction and MoE routing structure have been well verified by system \textit{LID\_SC} and \textit{MoE}. 

To further illustrate the effectiveness of \textit{LUPET}, several ablation studies are carried out to investigate the remained components.
By removing \textit{U} (acoustic unit discovery) from \textit{LUPET}, it can be clearly observed that WERs on high-resource languages consistently increase. Contrary to high-resource, some low-resource languages especially for tt, achieve somewhat improvements. It demonstrates the quality of discrete units that discovered by MLM is related to the amount of data. Hence, MLM can usually bring positive gains to high-resource languages. With a longer MLM effective period (Uto50ep), low-resource languages have obvious performance degradation. The gains towards high-resource gradually converge to the language en. When increasing the weight coefficient of $\mathcal{L}_{mlm}$ ($w_2 =1$), inferior results compared to the original \textit{LUPET} setting are presented for most of languages. 

Disabling both \textit{U} and \textit{P} (IPA sharing prediction) exhibits worse results.  Tab. 2 provides two comparison views to understand the independent effect of the component \textit{P}. (1) \textit{LUPET / LU} can be seen as ``\textit{MoE} + \textit{P}". Comparing it with \textit{MoE},  IPA sharing is found to be beneficial for all languages except tt. (2) When only removing \textit{P} from \textit{LUPET}, WERs on all languages basically degrade. Low-resource languages clearly give the worse results, where tr increase the absolute WER above 4\%.
One possible reason is that the independent \textit{U} would lead to information loss due to the masking mechanism in MLM, especially for low-resource languages. In LUPET, with the help of intermediate phoneme prediction (\textit{P}), the loss of information is largely mitigated, thus not presenting much worse results on low-resource languages.

\begin{table}[h]
\centering
\caption{Averaged WER (\%) of different systems by attention decoding. Note that Whisper-large-v2 decodes in a greedy manner, while other systems utilize beam search with beam\_size=20.} \vspace{-8pt}
\resizebox{0.8\linewidth}{!}{
\begin{tabular}{c|cccc}
\toprule
Model       & avg   & avg w/o tt & avg 5high & avg 5low \\ \midrule
Whisper     & 20.86 & 11.52      & 13.21     & 28.52    \\ \midrule
Mono        & 17.89 & 10.02      & 9.52      & 26.26    \\
Vanilla     & 10.43 & 8.78       & 8.88      & 11.99    \\
Oracle\_LID & 9.20  & 7.89       & 8.31      & \textbf{10.08}    \\
LID\_SC     & 10.22  & 8.62       & 8.62      & 11.83    \\
MoE         & 9.61  & 8.25       & 8.46      & 10.76    \\ \midrule
LUPET       & \textbf{9.15}  & \textbf{7.77}       & \textbf{7.95}      & 10.35    \\ \bottomrule
\end{tabular}}
\label{tab:attn_wer}
\vspace{-12pt}
\end{table}

\subsection{Results on Attention Decoding}
Tab.~\ref{tab:attn_wer} presents the averaged WER metrics of different systems by attention decoding. Not surprisingly, \textit{LUPET} exhibits the overall best performance, especially on high-resource languages, and significantly outperforms its CTC decoding counterpart by 3.94\% absolute WER. \textit{Whisper} is introduced as an external reference. As can be seen, the zero-shot performance of Whisper is easily surpassed even by \textit{Mono}, illustrating the importance of in-domain training. Furthermore, it is noted that the overall performance gaps between systems in attention decoding are far less than that of CTC decoding, e.g., comparing \textit{Oracle\_LID} and \textit{LUPET}. We hypothesize that the attention decoder served as a language model may help overfit the pattern in the specific domain. This also explains why attention decoding clearly outperforms CTC decoding in our experiments.

\section{Conclusions}
This paper presents a novel view to seamlessly incorporate hierarchical information into multilingual ASR. Multiple information in different granularity, i.e., LID, acoustic unit, phoneme, and token, form a path LUPET that unfolds with encoder layers.  Experiments carried out on 10 languages of Common Voice corpus illustrate the effectiveness of LUPET, even outperforming the system with oracle LID information. Different components in LUPET are proved to be useful in ablation studies. It is found that the acoustic unit discovery and phoneme prediction significantly help the recognition on high-resource languages, largely mitigating the compromised phenomenon.

\bibliographystyle{IEEEtran}
\bibliography{mybib}

\end{document}